\documentclass[12pt]{article}
\usepackage{graphicx}
\begin{document}
\rightline{NKU-2013-SF1}
\bigskip
\begin{center}
{\Large\bf Cold, ultracold and Nariai black holes with  quintessence}

\end{center}
\hspace{0.4cm}
\begin{center}
Sharmanthie Fernando \footnote{fernando@nku.edu}\\
{\small\it Department of Physics \& Geology}\\
{\small\it Northern Kentucky University}\\
{\small\it Highland Heights}\\
{\small\it Kentucky 41099}\\
{\small\it U.S.A.}\\

\end{center}

\begin{center}
{\bf Abstract}
\end{center}

\hspace{0.7cm} 

In this paper, we study the properties of the charged black hole surrounded by the quintessence.  The solution space for the horizons for various values of the mass $M$, charge $Q$, and the quintessence parameter $\alpha$ are studied in detail. Special focus in given to the degenerate horizons: we obtain cold, ultracold and Nariai black holes which has similar topologies as for the Reissner-Nordstrom-de Sitter black holes. We also study the lukewarm black hole with the quintessence in this paper.\\


\section{ Introduction}

There are strong observational support for the fact that universe is undergoing accelerated expansion leading to the presence of dark energy \cite{perl} \cite{reis}. It is one of the greatest challenges in modern physics to seek solutions as to why and how this acceleration occurs. Various dynamical models of dark energy have been considered in the literature to explain the acceleration \cite{edmund}. Most of these models involve a dynamical scalar field. In this paper, we investigate black holes surrounded by quintessence matter. Quintessence field is a scalar field coupled to gravity and is a candidate for dark energy. There are many works that have focused on the quintessence model. An extended quintessence model has been proposed by coupling the scalar field to the Ricci scalar in \cite{wang}. The correspondence between the quintessence and the tachyon dark energy has been studied in \cite{ave}. Dynamics of the phantom energy interacting with quintessence models are presented in \cite{jamil}. Local measurements that can detect the quintessence field has been suggested in \cite{romali}. Kiselev derived black hole solutions surrounded by the quintessence matter in an interesting paper \cite{kiselev}. In this paper, we will focus on studying the properties of charged black holes surrounded by the quintessence.

The paper is organized as follows. In section 2 we introduce the charged black hole surrounded by the quintessence. In section 3, we discuss the free quintessence model. In section 4,  the Reissnner-Nordstrom black hole with and without the cosmological constant is compared with the charged black hole with the quintessence. In section 5 and 6, the solution space of the horizons are discussed. In section 7, the charged Nariai and the cold black holes are discussed. In section 8, the ultra-cold black hole is presented. In section 9, the lukewarm black hole is given. Finally, in section 10, the conclusions are given.

\section { Charged  black hole surrounded by the quintessence}

In this section, we will give an introduction to the charged black hole surrounded by the quintessence, which was derived by Kiselev \cite{kiselev}. The geometry of the black hole is given by the metric,
\begin{equation}
ds^2 = - f(r) dt^2 + \frac{ dr^2}{ f(r)} + r^2 ( d \theta^2 + sin^2 \theta d \phi^2)
\end{equation}
Here,
\begin{equation}
f(r) = 1 - \frac{ 2 M} { r} + \frac{ Q^2}{ r^2} - 
\frac{ \alpha}{ r^{ 3 w_q + 1}}
\end{equation}
and $M$ is the mass, $Q$ is the charge, $ \alpha$ a normalization factor and $ w_q$ is the state parameter of the quintessence matter. In this paper, the parameter $w_q$ has the range,
\begin{equation}
-1 < w_q < - \frac{ 1}{ 3}
\end{equation}
The quintessence matter has the equation of state as,
\begin{equation}
p_q = w_q \rho_q
\end{equation}
and,
\begin{equation}
\rho_q = - \frac{ \alpha}{ 2} \frac { 3 w_q}{ r^{ 3 ( 1 + w_q)}}
\end{equation}
$p_q$ is the pressure and $\rho_q$ is the energy density of the matter. To cause acceleration, the pressure $p_q$ has to be negative, and the matter energy density $\rho_q$ is positive. Since $ w_q$ is negative, to obtain a positive pressure, the parameter $ \alpha$ has to be positive. The basis  for the choices of the parameters and for  details on the derivation of the metric, reader is referred to the original paper of Kiselev \cite{kiselev}.\\

For $ -1/3 \leq w_q < 1$, the solutions are asymptotically flat. For the range, 

\noindent
$ -1 < w_q < -1/3$, the space-time is non-asymptotically flat.\\

There are few works that have focused on studying this black hole. Thermodynamics and phase transitions has been studied in \cite{sal} for charged black holes for all possible values of $w_q$. In \cite{manu}, the authors studied the thermodynamics of only the asymptotically flat solutions and observed second order phase transitions. Quasinormal modes for a charged black hole surrounded by dark energy has been studied in \cite{var}. The null geodesics of the neutral black hole surrounded by the quintessence has been studied by Fernando in \cite{fernando}\\

When $ w_q = -1$, the function $f(r)$ reduces to,
\begin{equation}
f(r) = 1 - \frac{ 2 M} { r} + \frac{ Q^2}{ r^2} - \alpha r^2
\end{equation}
which is the Reissner-Nordstrom-de Sitter black hole (one can replace $\alpha = \frac{ \Lambda}{3}$ where $ \Lambda$ is the cosmological constant).

In the rest of the paper, we will choose $ w_q  = -\frac{ 2}{3}$ as the simplest, nontrivial charged black hole surrounded by the quintessence to study. This assumption will lead to the metric in consideration to be,
\begin{equation}
ds^2 = - f(r) dt^2 + \frac{ dr^2}{ f(r)} + r^2 ( d \theta^2 + sin^2 \theta d \phi^2)
\end{equation}
with,
\begin{equation}
f(r) = 1 - \frac{ 2 M} { r} + \frac{ Q^2}{ r^2} - \alpha r
\end{equation}
Hence, the black hole discussed in this paper is non-asymptotically flat and has a curvature singularity at $ r = 0$.

\section{ Free quintessence space-time with $M=0$ and $ Q =0$ }
When the mass and charge are zero, the geometry of the metric in eq.(7) simplifies to the free-quintessence space-time  given by,
\begin{equation}
ds^2 = -  ( 1 - \alpha r ) dt^2 + \frac{ dr^2}{ ( 1 - \alpha r)} + r^2 ( d \theta^2 + sin^2 \theta d \phi^2)
\end{equation}
Such a space-time has an outer horizon at $ r_c =\frac{ 1}{ \alpha}$. The space-time  has a curvature scalar $R = \frac{ 6 \alpha }{r}$. Hence there is a singularity at $ r =0$. This space-time is very similar to the de Sitter space-time which has the metric,

\begin{equation}
ds^2 = -  ( 1 - \frac{\Lambda r^2 }{3}) dt^2 + 
\frac{ dr^2}{ ( 1 - \frac{\Lambda r^2}{3})} + r^2 ( d \theta^2 + sin^2 \theta d \phi^2)
\end{equation}
with an outer horizon at $ r_c = \sqrt{ \frac{ 3}{ \Lambda}}$, which is the cosmological horizon. 

\begin{center}
\scalebox{.9}{\includegraphics{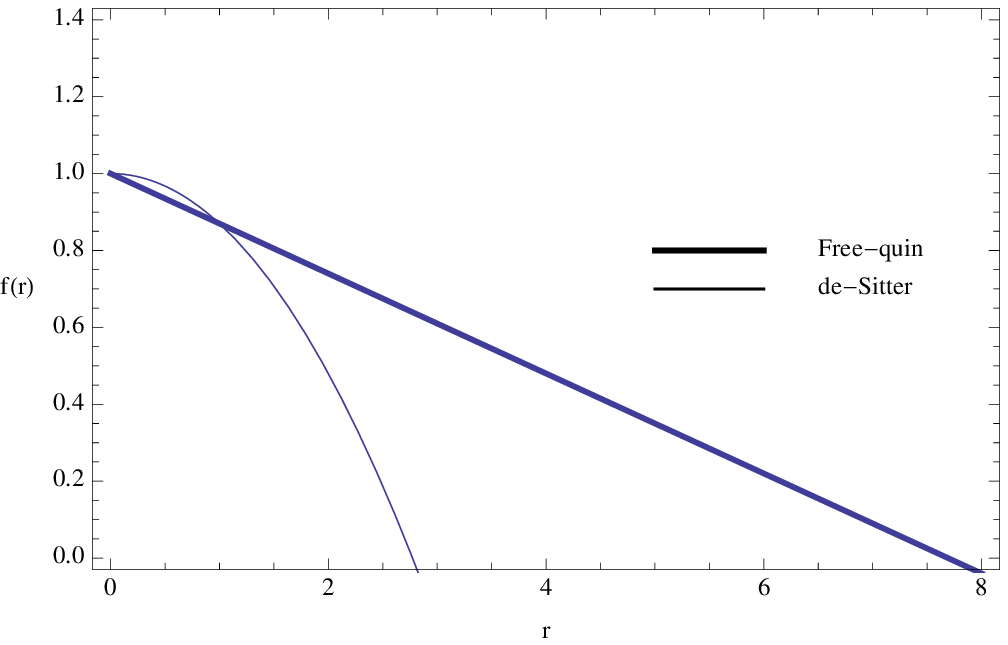}}

\vspace{0.3cm}

 \end{center}

Figure 1. The figure shows  $f(r)$ vs $r$ for free quintessence and the de Sitter space-time. Here $\alpha = \frac{\Lambda}{3} = 0.13$\\

\noindent
The space-time with the free quintessence has the Hawking temperature,
\begin{equation}
T_{H(free~quin)} = \frac{ 1}{ 4 \pi} \left | \frac{d g_{tt}}{dr} \right| _{ r = r_h}= \frac{ \alpha}{ 4 \pi}
\end{equation}
The de Sitter cosmology has the background temperature as,
\begin{equation}
T_{H(de~Sitter)} = \frac{ 1 }{ 2 \pi} \sqrt{ \frac{\Lambda} { 3} }
\end{equation}
A detailed comparison of the free-quintessence space-time and the de Sitter space-time is presented in \cite{kiselev}. A description of the de-Sitter space-time can be found in the book by Griffiths and Podolsk$\acute{y}$ \cite{pod}.

\section{ Reissner-Nordstrom black hole with and without  the cosmological constant}

In order to fully understand and appreciate the properties of the charged black holes with the quintessence, we will review properties of the  Reissner-Nordstrom black hole and the Reissner-Nordstrom-de Sitter black hole in this section.\\

The Reissner-Nordstrom black hole is given by the metric,

\begin{equation}
f(r) = 1 - \frac{ 2 M}{r} + \frac{ Q^2} {r^2}
\end{equation}
There are two horizons,
\begin{equation}
r_+  = M - \sqrt{ M^2 - Q^2}; \hspace{0.4 cm} r_{++} = M + \sqrt{ M^2 - Q^2}
\end{equation}
Black hole exists only when $ M \geq Q$. Hence the $Q_{critical} = M$. When $ Q= Q_{critical}$, there is a degenerate horizon. When $ Q < Q_{critical}$, there are two horizons. When $ Q > Q_{critical}$, there are no horizons and the solution becomes a naked singularity. 
\\

Since both the cosmological constant and the quintessence matter provide a mechanism for the acceleration of the universe, it is important to compare the charged black hole in de-Sitter space and with the quintessence matter. The Reissner-Nordstrom-de Sitter black hole has the metric,
\begin{equation}
ds^2 = - f(r) dt^2 + \frac{ dr^2}{ f(r)} + r^2 ( d \theta^2 + sin^2 \theta d \phi^2)
\end{equation}
with,
\begin{equation}
f(r) = 1 - \frac{ 2 M} { r} + \frac{ Q^2}{ r^2} - 
\frac{ \Lambda}{3} r^2
\end{equation}

\begin{center}
\scalebox{.9}{\includegraphics{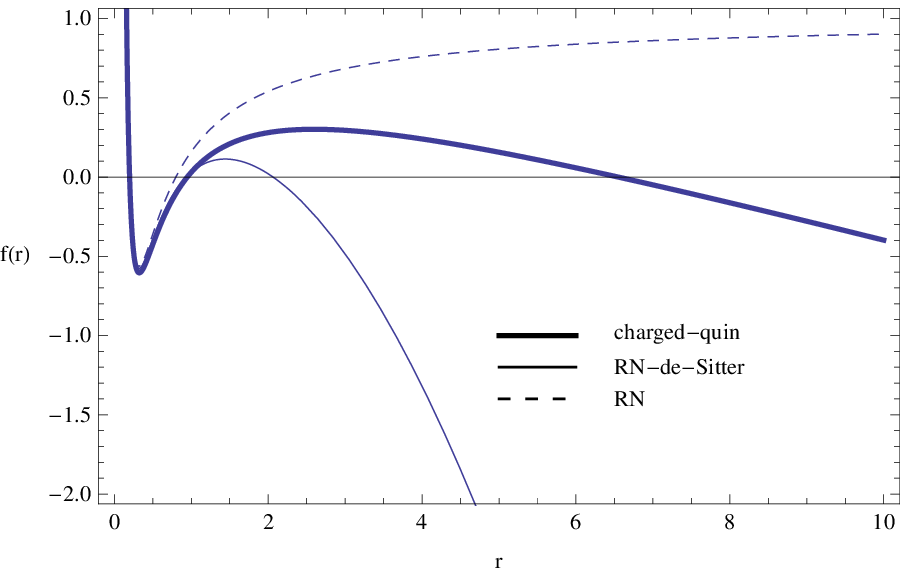}}

\vspace{0.3cm}

 \end{center}

Figure 2. The figure shows  $f(r)$ vs $r$ for charged black hole with the  quintessence, Reissner-Nordstrom black hole
 and  and the Reissner-Nordstrom-de Sitter space-time. Here $M=0.5, Q = 0.4$ and $\alpha = \frac{\Lambda}{3} = 0.13$\\

Reissner-Nordstrom-de Sitter black hole also has the possibility of having three horizons for the appropriate values of $M$, $Q$ and $\Lambda$: inner black hole horizon ($r_+$), outer black hole horizon($r_{++}$) and the cosmological horizon ($r_c$). The quantum global structure of the de Sitter space and the charged black holes in de Sitter space is discussed in detail in \cite{raphael}. The two horizons, $r_{++}$ and $r_c$, are not in thermal equilibrium. However, there are two families of the Reissner-Nordstrom-de-Sitter black hole in which the temperatures at the event horizon and the cosmological horizon are the same: charged Nariai black hole and the lukewarm black hole. Charged Nariai black hole is a solution where the cosmological horizon and the black hole horizon coincides.  Even though the horizons seem to coincide, the proper distance between them are non-zero. Also, the temperature is non-zero and both horizons have the same temperature. A detailed description of the thermodynamics of the Reissner-Nordstrom-de Sitter black hole is given in \cite{mann2}\cite{piazzo}\cite{romans}. Instantons in the Reissner-Nordstrom-de Sitter space is discussed in \cite{ross}\cite{mann}. Generalized Nariai solutions for Yang-type monopoles are discussed in \cite{diaz}. Charged  Nariai black hole with a dilaton is discussed in \cite{raphael3}.\\

The temperature of the Reissner-Nordstrom-de-Sitter black hole is given by,
\begin{equation}
T_H = \frac{ 1} { 4 \pi r_{++}} \left| 1 - \frac{ Q^2}{r_{++}^2} - 3 \Lambda r_{++}^2 \right|
\end{equation}

The temperature of a general  charged black hole with the quintessence black hole is given by,
\begin{equation}
T_H = \frac{ 1} { 4 \pi r_{++}} \left| 1 - \frac{ Q^2}{r_{++}^2} - 2 \alpha r_{++}\right|
\end{equation}

\begin{center}
\scalebox{.9}{\includegraphics{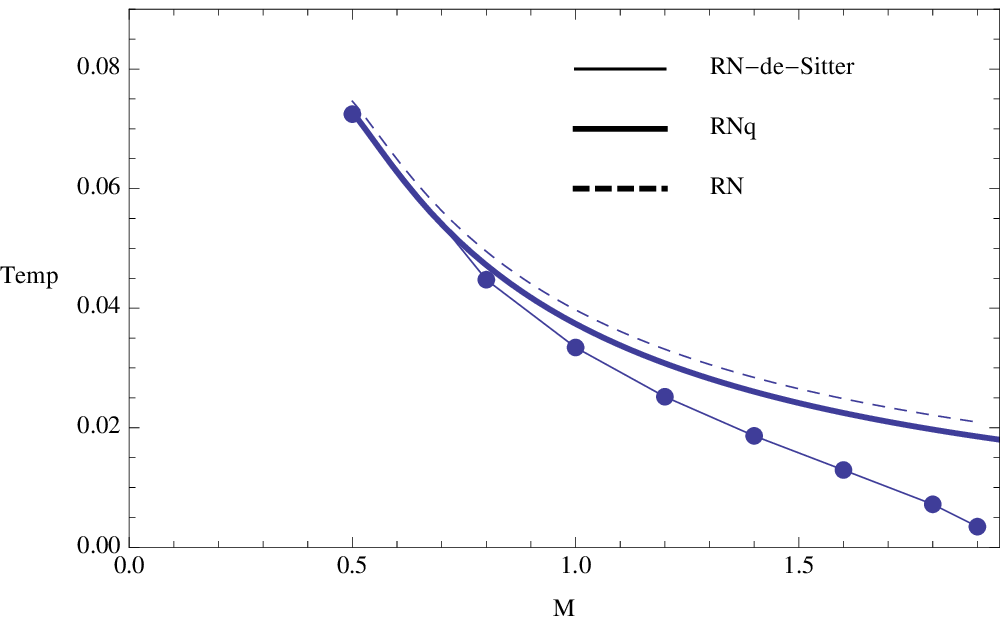}}

\vspace{0.3cm}
\end{center}
Figure 3. The figure shows the graphs Temperature vs $M$ for the charged black holes with the cosmological constant, without the cosmological constant, and with the quintessence matter. Here, $Q = 0.4$ and $ \alpha = \frac{\Lambda}{3} = 0.01$.\\

\noindent
From the graph in Fig.3, it is clear that Reissner-Nordstrom-de Sitter black hole is colder than the others.

\section{ The solution space for  the horizons of the charged black hole with the quintessence}

The horizons for the black holes are obtained from the roots of the function $f(r)=0$ given by,
\begin{equation}
\alpha r^3 - r^2 + 2 M r - Q^2 =0
\end{equation}
which is a cubic equation. We use well known results on the roots of a cubic polynomial  here. The discriminant $\bigtriangleup$   for the cubic equation in eq.(19) is given by,
\begin{equation}
\bigtriangleup = 4 (M^2 - Q^2 ) + \alpha ( - 32 M^3 + 36 Q^2 M ) - 27 \alpha^2 Q^4
\end{equation}
Note that when $ \alpha =0$, the discriminant $ \bigtriangleup = 4 ( M^2 - Q^2)$ which indicates the nature of the roots of the Reissner-Nordstrom black hole. Depending on the sign of $\bigtriangleup$, there would be one horizon(degenerate case), two horizons  or no horizons. 

Now getting back to the case for $\alpha \neq 0$, we will analyze all the possibilities in the following sections.

\subsection{ Structure of roots for fixed M}

Now, one can study the behavior of $\bigtriangleup$ for fixed $M$ and varying $Q$. One can solve $\bigtriangleup =0$ for $Q_{critical}$ in terms of $M$ and $\alpha$. Since $ \bigtriangleup$ is function of $Q^2$, it can be easily solved to be
\begin{equation}
Q^2_{critical (1,2)} = \frac{ 2 \left( -1 + 9 \alpha M \pm \sqrt{ 1 - 18 \alpha M + 108 \alpha^2 M^2 - 216 \alpha^3 M^3 } \right)}{ 27 \alpha^2}
\end{equation}
$Q^2_{critical(1,2)}$ is plotted in the Fig.4. One can observe that $Q^2_{critical1} >0$ for all values of $M$ while $Q^2_{critical2} >0$ only for range of values of $M$.

\begin{center}
\scalebox{.9}{\includegraphics{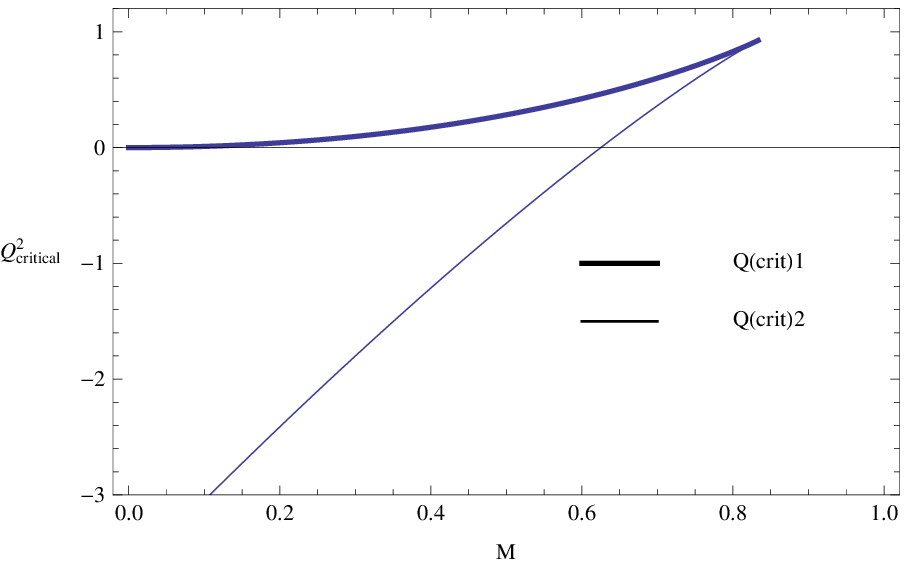}}

\vspace{0.3cm}
\end{center}

Figure 4. The figure shows the graphs for $Q^2_ {critical (1,2)}$ vs $M$. Here $\alpha = 0.2$. \\

\noindent
In Fig.5 and Fig.6, the $Q_{critical}$ is plotted with $M$. The function $f(r)$ is  plotted for one particular value of $M$ and the corresponding $Q_{critical}$. It is clear  that the location of the degenerate root differs for $Q_{critical1}$ and $Q_{critical2}$.

\begin{center}
\scalebox{.9}{\includegraphics{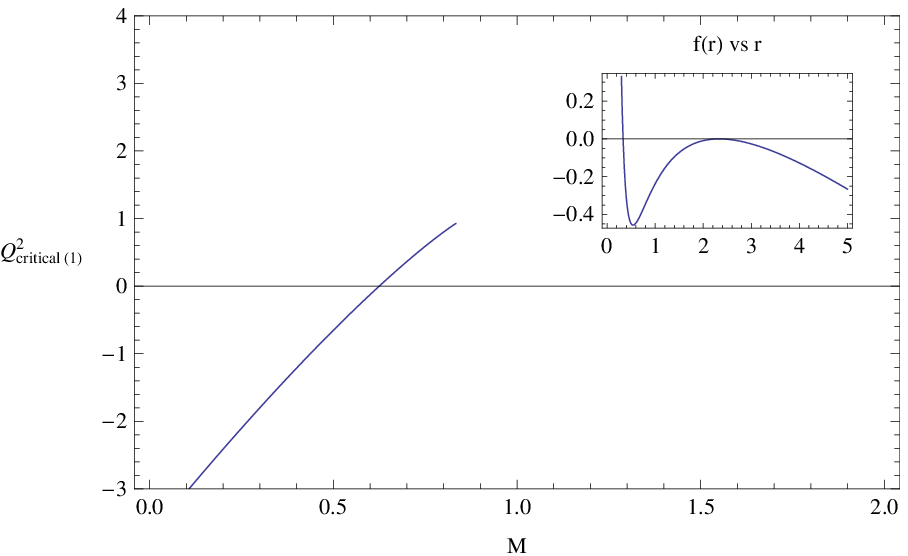}}

\vspace{0.3cm}

\end{center}

Figure 5. The figure shows the $Q_{critical}1$ vs $M$. The function $f(r)$ is plotted for $M = 0.7$ and $Q_{critical1}= 0.6025$. Here $\alpha = 0.2$.

\begin{center}
\scalebox{.9}{\includegraphics{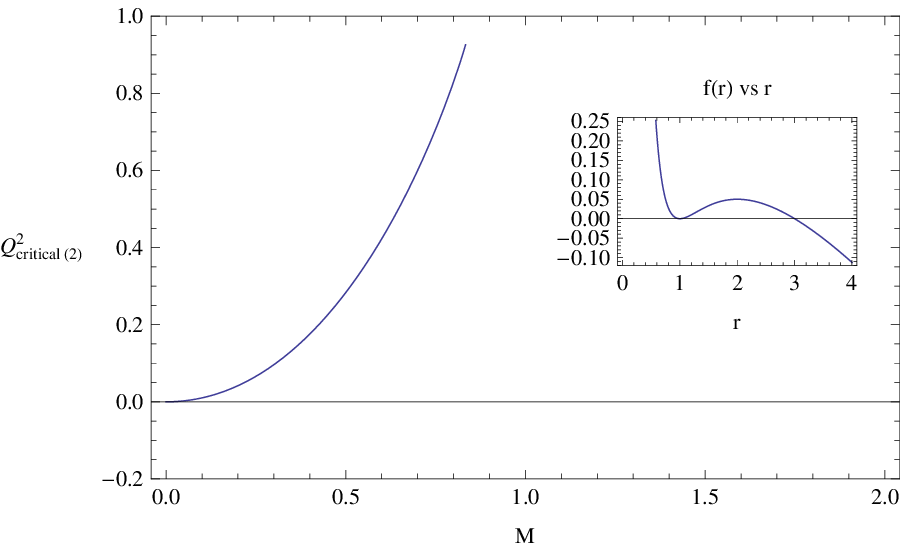}}

\vspace{0.3cm}

\end{center}

Figure 6. The figure shows the $Q_{critical2}$ vs $M$. The function $f(r)$ is plotted for $M = 0.7$ and $Q_{critical2}= 0.7746$. Here $\alpha = 0.2$.\\

\noindent
Now that we have understood  $\bigtriangleup =0$ case well, one can study what type of roots are given for $f(r)=0$ for non-zero values of $\bigtriangleup$. The graph Fig.7 shows how $\bigtriangleup$ changes with $Q$ and the corresponding possibilities for the roots of $f(r)=0$. The corresponding behavior for $f(r)$ is also plotted underneath it. Each possibility gives different scenarios for the horizons.

\begin{center}
\scalebox{.9}{\includegraphics{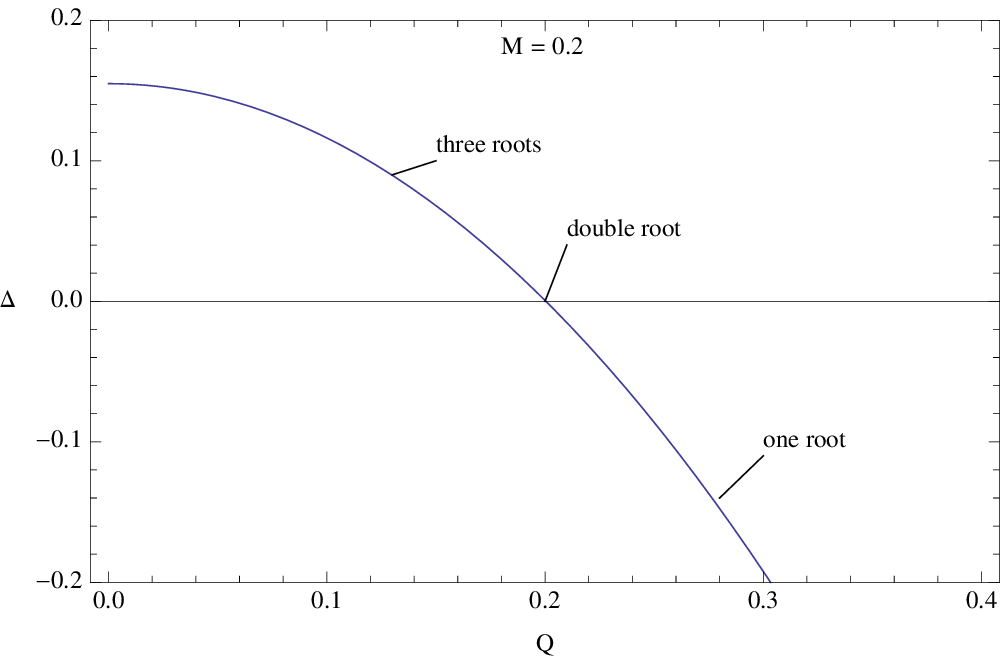}}

\vspace{0.3cm}

\scalebox{.9}{\includegraphics{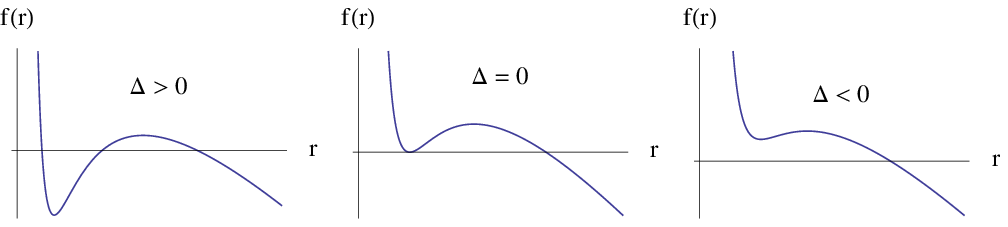}}

\end{center}

Figure 7. The figure shows the $\bigtriangleup$ vs $Q$ and the corresponding graphs $f(r)$ vs $r$. Here $Q = 1$ and $\alpha = 0.05$. \\

In the Fig.8, the function $f(r)$ is plotted for three values of $Q$ with fixed $M$. 

\begin{center}

\scalebox{.9}{\includegraphics{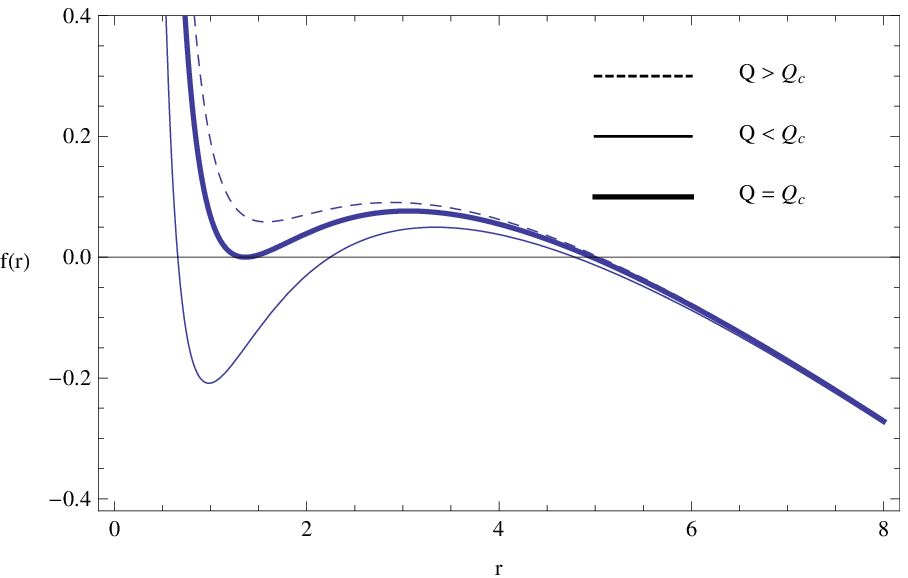}}

\end{center}

Figure 8. The figure shows the $f(r)$ vs $r$  for various values of $Q$.\\

\subsection{ Structure of roots for fixed Q}

Now, one can study the behavior of $\bigtriangleup$ for fixed $Q$ and varying $M$. One can solve $\bigtriangleup =0$ for $M_{critical}$ in terms of $Q$ and $\alpha$. Since $\bigtriangleup$ is a cubic polynomial in $M$, there will be three roots for $M_{critical}$ (given $Q$ and $\alpha$). The solution is quite lengthy and we will avoid writing the expressions explicitly. The three solutions for $M_{critical}$ are plotted in Fig.9. The first and the second  are realistic values since the other is negative.

\begin{center}
\scalebox{.9}{\includegraphics{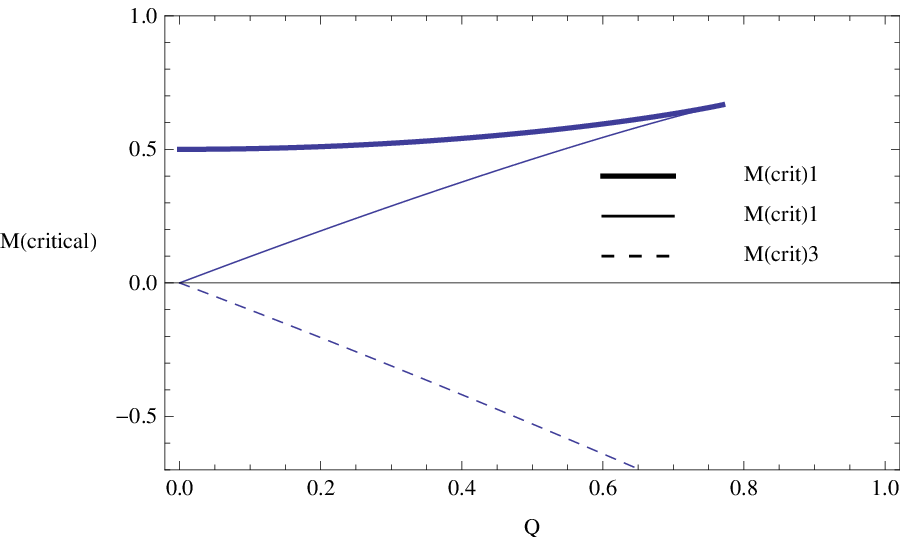}}

\vspace{0.3cm}

\end{center}

Figure 9. The figure shows the $M_{critical}$ vs $Q$. The three roots are given in the three graphs. Here,  $\alpha = 0.25$. \\

In Fig.10 and Fig.11, the $M_{critical}$ is plotted with $Q$. The function $f(r)$ is  plotted for one particular value of $Q$ and the corresponding $M_{critical}$,  It is clear  that the location of the degenerate root differs for $M_{critical1}$ and $M_{critical2}$.

\begin{center}
\scalebox{.9}{\includegraphics{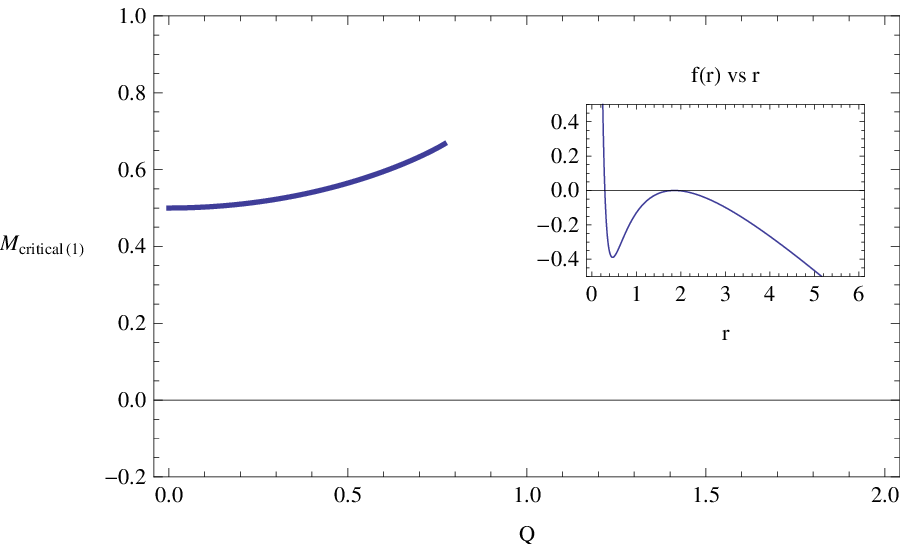}}

\vspace{0.3cm}

\end{center}

Figure 10. The figure shows the $M_{critical}1$ vs $Q$. The function $f(r)$ is plotted for $Q = 0.5$ and $M_{critical1}= 0.5648$. Here $\alpha = 0.25$.\\

\begin{center}
\scalebox{.9}{\includegraphics{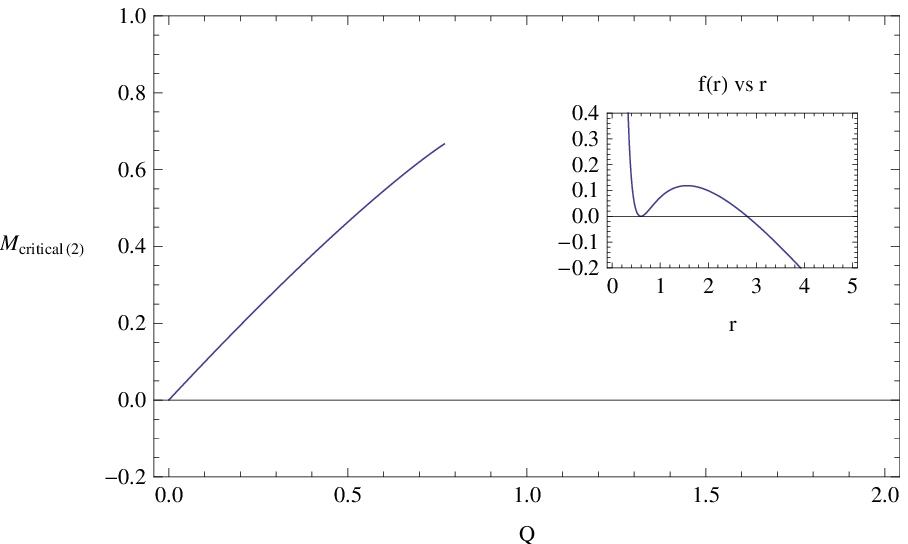}}

\vspace{0.3cm}

\end{center}

Figure 11. The figure shows the $M_{critical2}$ vs $Q$. The function $f(r)$ is plotted for $Q = 0.5$ and $M_{critical2}= 0.4633$. Here $\alpha = 0.25$.\\

Now that we have understood how $\bigtriangleup =0$ case well, one can study what type of roots are given for $f(r)=0$ for non-zero values of $\bigtriangleup$. The graph in Fig.12 shows how $\bigtriangleup$ changes with $M$ and the corresponding possibilities for the roots of $f(r)=0$. The corresponding behavior for $f(r)$ is also plotted underneath it. Each possibility gives different scenarios for the horizons.

 \newpage

\begin{center}
\scalebox{.9}{\includegraphics{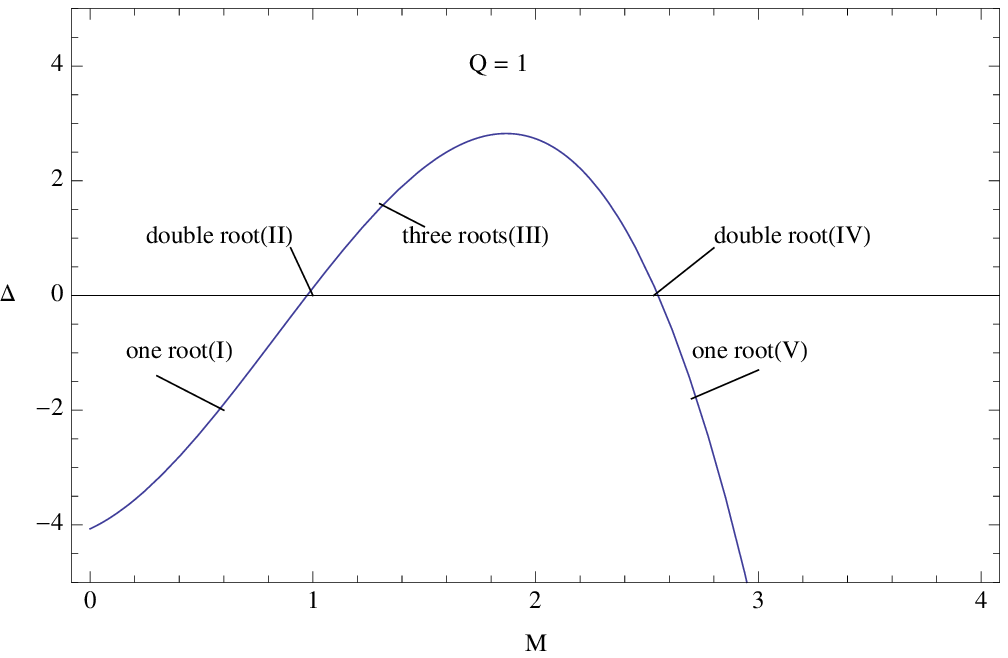}}

\vspace{0.3cm}

\scalebox{.9}{\includegraphics{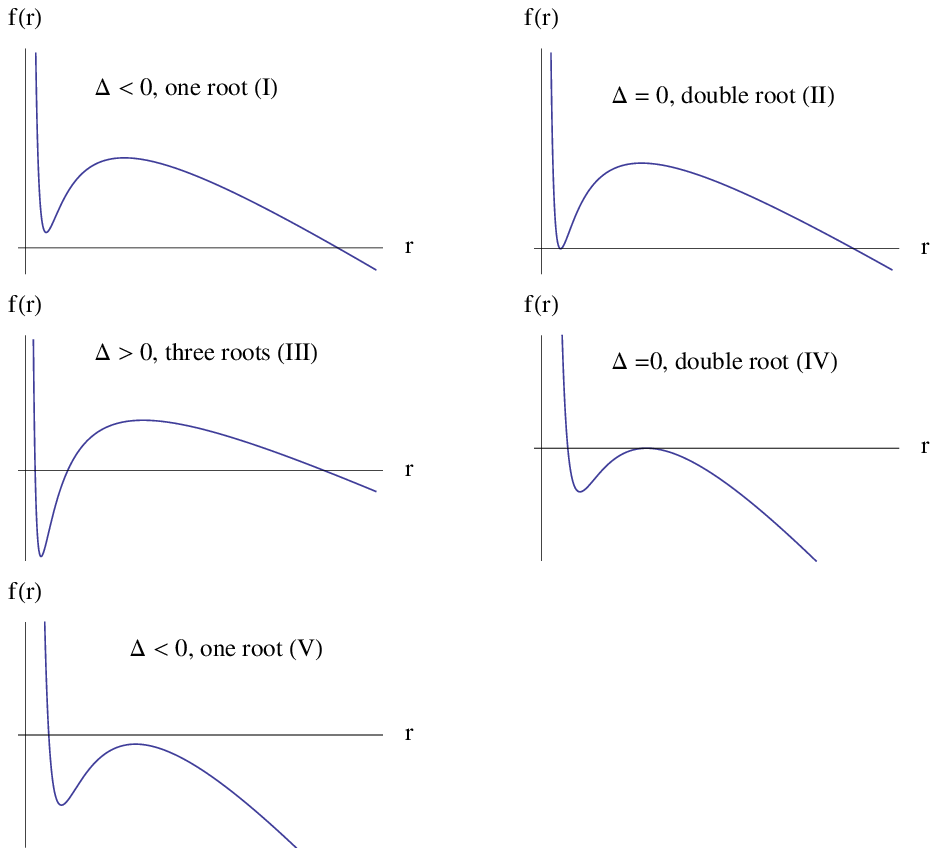}}

\end{center}

Figure 12. The figure shows the $\bigtriangleup$ vs $M$ and the corresponding graphs $f(r)$ vs $r$. Here $Q = 1$ and $\alpha = 0.05$. \\

\noindent
To summarize, for $ M = M_{critical}$, there will be a degenerate horizon where two out of the three horizons will coincide. When $ M > M_{critical}$, there will be only one horizon. When $ M < M_{critical}$, there will be three horizons, the Cauchy horizon ($r_+$), the event horizon($r_{++}$) and the cosmological horizon ($r_c$). This is represented for one particular case in Fig.13.

\begin{center}
\scalebox{.9}{\includegraphics{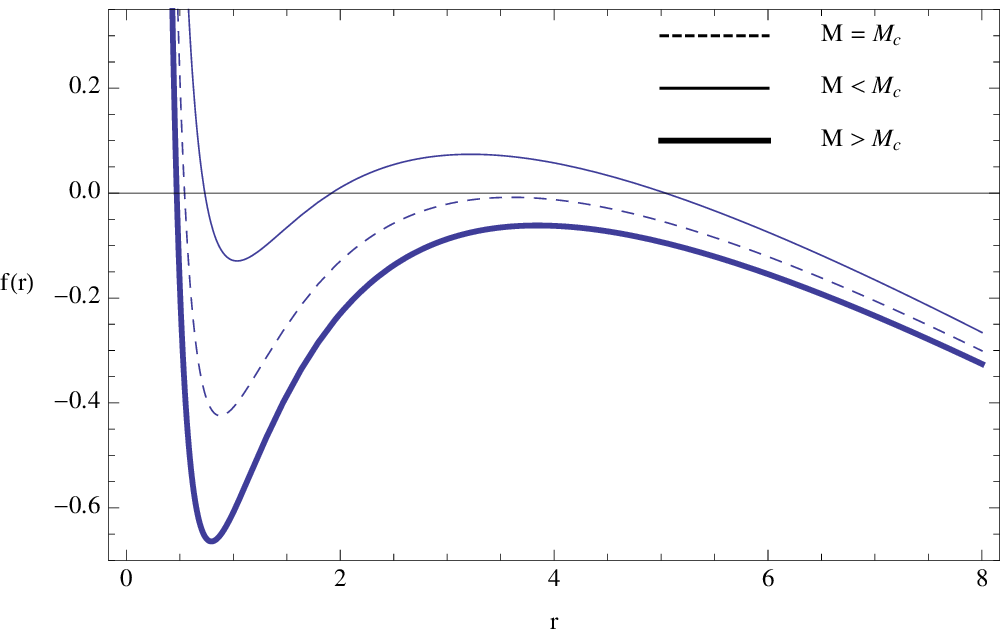}}

\vspace{0.3cm}
\end{center}

Figure 13. The figure shows the graphs for $f(r)$ vs $r$ for fixed charge and varying mass. Here $Q= 0.96$ and $\alpha = 0.13$. \\

\subsection{ Solution space}

Summarizing all possibilities for  horizons, one can draw the graph in Fig.14 where M is plotted against Q. When $Q=0$, one get free-quintessence, Neutral-quintessence and the Neutral-Nariai black hole. For other values of $Q$, when the points are inside the boundaries, one have the general charged-quintessence black hole with three horizons. On the boundaries, for $M = M_{critical1}$ one obtain charged Nariai black hole with the quintessence. The properties of such black holes will be described in detail in section(7). When $M = M_{critical2}$, one obtain cold black holes with the quintessence. They are described in detail in section(7). When both boundaries meet, one obtain the ultra cold black holes which have zero temperature. They are described in section(8).

\begin{center}

\scalebox{.9}{\includegraphics{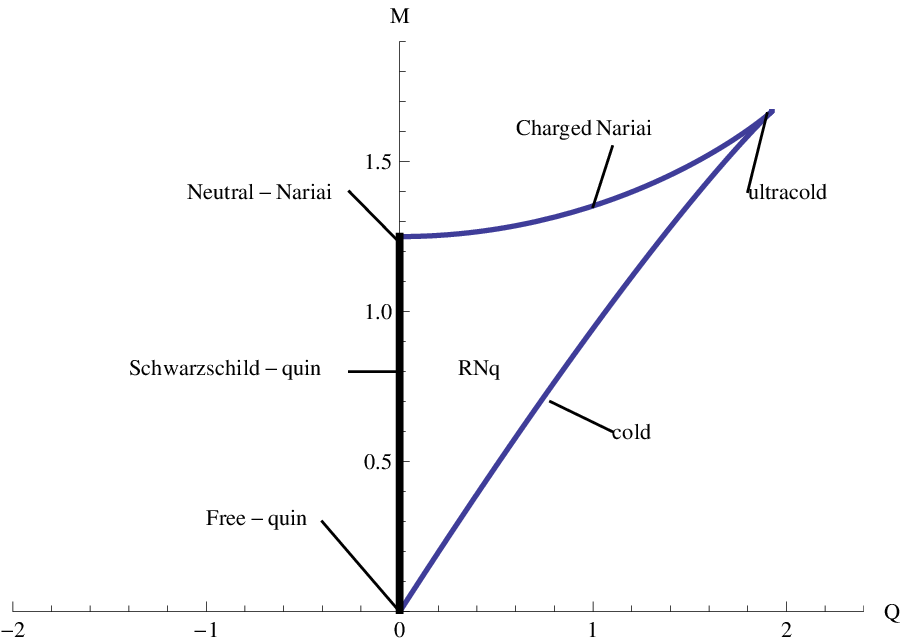}}

\end{center}

Figure 14. The figure shows  $M$ vs $Q$ graph. Here, $\alpha = 0.1$. \\

\section{ Number of horizons for the charged black hole with the quintessence}

Depending on the values of $M, Q$ and $\alpha$, the function $f(r)$ could have one, two or three roots leading to different kind of black holes. Here, we will explicitly write down the horizon radius for all three different scenarios.

\subsection{ The black hole with three horizons}

When $\bigtriangleup >0$  there are three roots to the function $f(r)$. The smallest root  corresponds to the black hole Cauchy horizon ($r_+$), the second one corresponds to the black hole event horizon ($r_{++}$)  and the largest root corresponds to the cosmological horizon ($r_c$). Then, the three horizons are given by,
\begin{equation}
r_c = 2 \sqrt{\frac{-p}{3}} cos\left( \frac{\theta}{3}\right) + \frac{ 1}{ 3 \alpha}
\end{equation}
\begin{equation}
r_{++} = 2 \sqrt{\frac{-p}{3}} cos\left( \frac{\theta}{3} - \frac{ 2 \pi}{3}\right) + \frac{ 1}{ 3 \alpha}
\end{equation}
\begin{equation}
r_{+} = 2 \sqrt{\frac{-p}{3}} cos\left( \frac{\theta}{3} - \frac{  \pi}{3}\right) + \frac{ 1}{ 3 \alpha}
\end{equation}
Here the three functions $p$, $q$ and $\theta$ are,
\begin{equation}
p = \frac{ ( 6 M \alpha - 1)}{ 3 \alpha^2}
\end{equation}
\begin{equation}
q = \frac{ (  - 2 + 18 \alpha M - 27 \alpha^2 Q^2)}{ 27 \alpha^3}
\end{equation}

\begin{equation}
\theta = cos^{-1}\left( \frac{ 3 q}{ 2 p} \sqrt{ \frac{ - 3}{ p} }\right)
\end{equation}

\noindent
Notice that for small $\alpha$, the cosmological horizon $r_c$ $\approx$ $\frac{ 1}{\alpha}$ and $M \ll \frac{1}{\alpha}$. For the Reissner-Nordstrom-de-Sitter black hole, for special values of $M$, $Q$ and the cosmological constant $\Lambda$, the possibility of three horizons exists \cite{romans}. The horizon at $r_c$ here is similar to the cosmological horizon in the Reissner-Nordstrom-de-Sitter case.
 
\subsection{Black hole with two horizons}

In this case, there will be a double root and another simple root for the equation, $f(r) =0$ as,
\begin{equation}
r_1 = r_2 = \frac{ (9 \alpha Q^2 - 2 M)}{ 2 ( -1 + 6 M \alpha)}
\end{equation}
and,
\begin{equation}
r_3 = \frac{  (-1 + 8 \alpha M - 9 \alpha^2 Q^2)}{ \alpha ( -1 + 6 M \alpha)}
\end{equation}

\subsection{ Black hole with one horizon}

There are two possibilities to have only one horizon. One, is to have a triple root for $f(r)=0$. The other is to have one real root and two complex conjugate roots for $f(r)=0$.

\subsubsection{ Triple root }

The triple real root given by,
\begin{equation}
r_{1} = r_{1} = r_3 = \frac{ 1}{ 3 \alpha}
\end{equation}

\subsubsection{ Black hole with one horizon and non zero temperature}

In this case,  there are two separate scenarios.\\

{\bf Case 1:  $ M < \frac{ 1 }{ 6 \alpha}$, ($p <0$)}

\begin{equation}
\eta = \frac{ -q}{2} \left( \frac{ 3 }{ |p| } \right)^{(3/2)}
\end{equation}
\noindent
Here, the real root is given by,
\begin{equation}
x_{real} =  \frac{ |\eta|}{\eta} \sqrt{ \frac{ 4 |p|}{3} } 
cosh \left( \frac{ 1}{3} cosh^{-1} (|\eta|)\right) - \frac{ 1}{ 3 \alpha}
\end{equation}

{\bf case 2: $ M > \frac{ 1}{ 6 \alpha}$ ($p > 0$)}\\

\noindent
Here, the real root is given by,
\begin{equation}
x_{real} =   \sqrt{ \frac{ 4 |p|}{3} } 
sinh \left( \frac{ 1}{3} sinh^{-1} (|\eta|)\right) - \frac{ 1}{ 3 \alpha}
\end{equation}
Note that $p$ and $q$ are given by eq.(25) and eq.(26).

\section{ Charged Nariai and cold black holes}

According to the description in section(5.2), when $\bigtriangleup =0$ and $(-1 + 6 M \alpha)  \neq 0$, there are double roots and a simple root for the equation $ f(r)=0$.    If the double roots occur at $ r = \rho$, then,
\begin{equation}
f(\rho)=0; \hspace{0.7 cm} f'(\rho) = 0
\end{equation}
By combining those two, one can obtain the following expressions for the charge $Q$ and the mass $M$ as,
\begin{equation}
M = \frac{\rho}{2} ( 2 - 3 \alpha \rho)
\end{equation}
\begin{equation}
Q^2 = \rho^2 ( 1 - 2 \alpha \rho)
\end{equation}
By substituting the above value of $M$ and $Q$ and factorizing the function $f(r)$, it can be rewritten as,
\begin{equation}
f(r) = \frac{ ( r - \rho)^2 \left( 1 - \alpha( 2 \rho + r) \right)}{r^2}
\end{equation}
There is another positive real root $b$ for the function $f(r)$ given by,
\begin{equation}
 b = \frac{1}{\alpha} - 2 \rho
\end{equation}
which is derived from the function in eq.(37).
Now, one can write $f(r)$, $M$, $Q$, and $\alpha$ in terms of $b$ and $\rho$ as,
\begin{equation}
f(r)_{Nariai/cold}(r) = \frac{ - ( r - b) ( r - \rho)^2}{ r^2 ( b + 2 \rho)}
\end{equation}

\begin{equation}
\alpha = \frac{ 1}{ ( b + 2 \rho)};\hspace{0.4 cm} M = \frac{ \rho( 2 b + \rho)}{ 2 ( b + 2 \rho)};\hspace{0.4 cm} Q^2 = \frac{ b \rho^2}{ b + 2 \rho}
\end{equation}
From the section(5.2), the degenerate root $\rho$ and the other simple root $b$ can be written explicitly in terms of $M$ and $Q$ as,
\begin{equation}
\rho = \frac{ - 2 M + 9 \alpha Q^2)}{ 2 ( - 1 + 6 \alpha M)}
\end{equation}
\begin{equation}
b = \frac{ (- 1 + 8 \alpha M - 9 \alpha^2 Q^2)}{\alpha ( - 1 + 6 \alpha M)}
\end{equation}
There are two possibilities for $b$: either it is outside the degenerate horizon $\rho$, or, it is inside the degenerate horizon. Which horizon is bigger depends on the expression,
\begin{equation}
\eta = -2 + 18 \alpha M - 27 \alpha^2 Q^2
\end{equation}
When $\eta < 0$ $\Rightarrow$ $ \rho < b$ and when $\eta > 0$ $\Rightarrow$ $ \rho > b $.\\

The Hawking temperature  at the horizon $b$   is,
\begin{equation}
T_H ( b) = \frac{ ( 1 - \frac{\rho}{b})^2}{ 4 \pi ( b + 2 \rho) }
\end{equation}

Since $ \alpha = \frac{ 1}{ b + 2 \rho}$,
\begin{equation}
T_H(b) = 
\frac{ \alpha}{ 4 \pi}  \left( 1 - \frac{ \rho}{ b} \right)^2
\end{equation}\\

\noindent
{\bf Case 1 ( $b > \rho$): Cold black hole}
\\

\noindent
When $b > \rho$, the black hole is called the  cold black hole and,
\begin{equation}
b > \rho \Rightarrow 0 < \rho < \frac{ 1}{ 3 \alpha}
\end{equation}
Also, the temperature at the black hole event horizon, $\rho$ is zero.

\begin{center}
\scalebox{.9}{\includegraphics{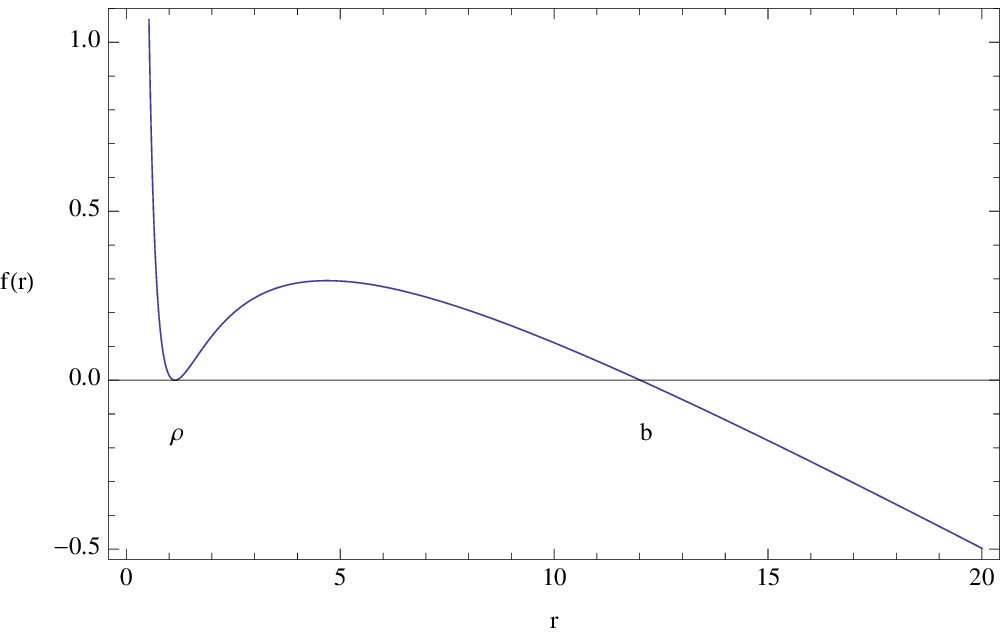}}

\vspace{0.3cm}
\end{center}

Figure 15. The figure shows the graphs for $f(r)$ vs $r$ for the cold black hole.  Here $Q= 1.04121$ , $ M = 1$ and $\alpha = 0.07$. \\

\noindent
If $ b > \rho $, $\Rightarrow$ $ 1 - \frac{\rho}{b} < 1$ which leads to,
\begin{equation}
T_H(b) < \frac{ \alpha}{ 4 \pi} = T_H(free-quintessence)
\end{equation}
Hence the temperature of the cosmological horizon is smaller due to the presence of the black hole compared to the free-quintessence case.\\
\noindent

To understand the geometry near the degenerate horizon, we can choose a new coordinate $y$ such as\cite{nam}
\begin{equation}
r = \rho - \epsilon y
\end{equation}
The function $f(r)$ can be expanded around $r = \rho$ as,
\begin{equation}
f(r) \approx  \frac{f''(\rho)}{2} ( \epsilon y)^2
\end{equation}
Also, let a new time coordinate be defined as $ \psi = \epsilon t $.  With these transformations, the metric is approximated to be,
\begin{equation}
ds^2 = \frac{ -f''(\rho)}{2} y^2 d \psi ^2 + \frac{ 2}{ f''(\rho)} \frac{ dy^2}{y^2}  + \rho^2  d \Omega^2
\end{equation}
Note that $ f''(\rho) >0$ for the cold black hole. The above geometry represents $AdS_2 \times S^2$ topology. The $AdS_2$ has the curvature $f''(\rho)/2$. The  cold Reissner-Nordstrom-de Sitter black hole has the same topology but with curvature  $\Lambda$.\\

{\bf Case 2 ( $b < \rho$ ):  Charged Nariai black hole}\\
\noindent
\begin{equation}
b < \rho \Rightarrow \frac{1}{ 3 \alpha} < \rho < \frac{ 1}{ 2 \alpha}
\end{equation}
\begin{center}
\scalebox{.9}{\includegraphics{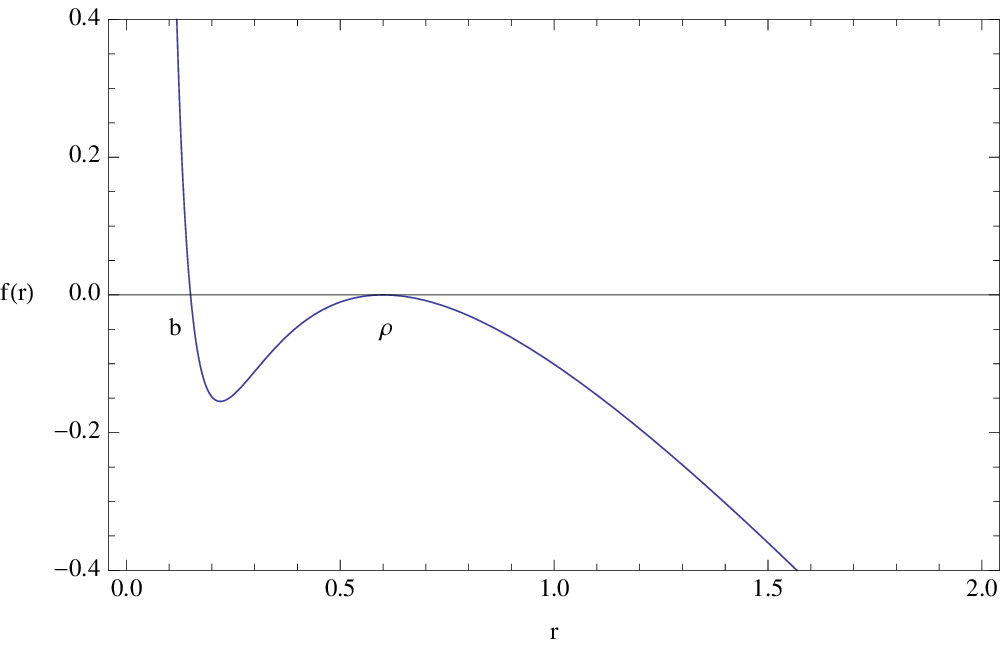}}

\vspace{0.3cm}
\end{center}

Figure 16. The figure shows the graphs for $f(r)$ vs $r$ for the charged Nariai black hole.  Here $Q= 0.2$ , $ M = 0.2$ and $\alpha = 0.740741$. \\

To find the geometry close to the degenerate horizon, one can approximate $f(r)$ by a parabola \cite{pod},
\begin{equation}
f(r) = \frac{ f''(\rho)}{2} ( r - r_1) ( r - r_2)
\end{equation}
Here, $r_1$ and $r_2$ represents a pair of close horizons, $r_{++}, r_c$. One can introduce new coordinates as,
\begin{equation}
t = \frac{ 2 \psi}{ \epsilon f''(\rho) };  ~ ~ r = \rho + \epsilon cos \chi
\end{equation}
$\chi=0$ corresponds to $r_1$ and $\chi = \pi$ corresponds to $r_2$. Substituting these new coordinates to the metric will yield,
\begin{equation}
ds^2 = \frac{ -2}{ f''(\rho)} \left( - sin^2\chi d \psi^2 + d \chi^2 \right) + \rho^2 d \Omega^2
\end{equation}
Note that $ f''(\rho) < 0$ for the Nariai black hole. Now, the above geometry corresponds to $ dS_2 \times S^2$. The curvature $ \Lambda_{eff}$  for $dS_2$ is given by  $ |f''(\rho)|/2$.  Near the degenerate horizon Reissner-Nordstrom-de Sitter black hole also has the same topology but with the curvature, $\Lambda$.

\section{ Ultra-cold black holes}

A special case occurs when $b$ and $ \rho$ coincides leading to a triple root for $f(r)=0$. As discussed in section(6.3.1), the triple real root is given by,
\begin{equation}
r_{+} = r_{++} = r_c = \frac{ 1}{ 3 \alpha}
\end{equation}
where,
\begin{equation}
 M = \frac{ 1 }{ 6 \alpha}; \hspace{0.7 cm} Q = \frac{ 1}{ 3 \sqrt{3} \alpha}
\end{equation}

\begin{center}
\scalebox{.9}{\includegraphics{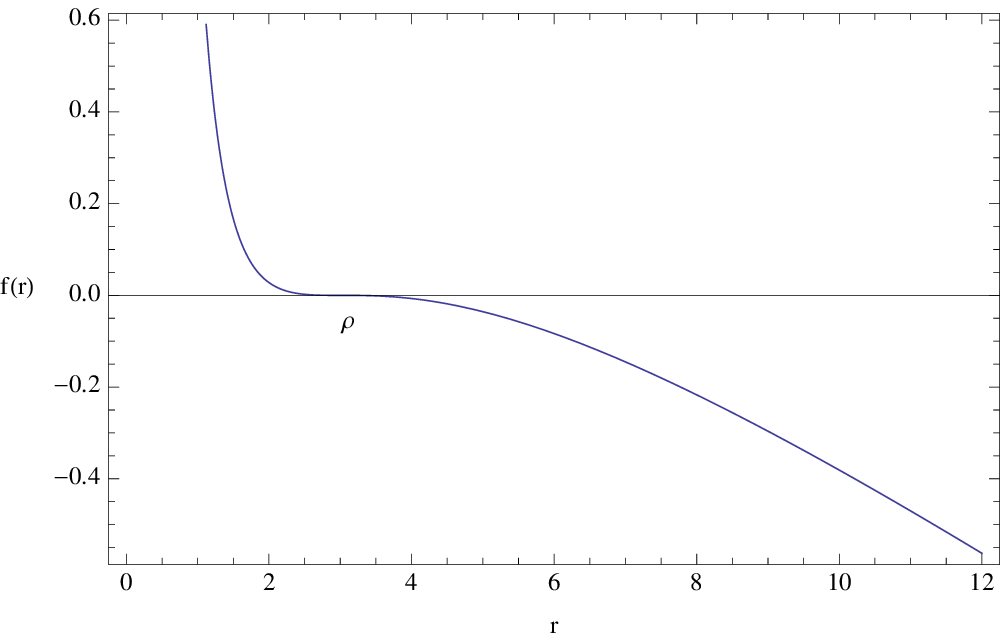}}

\vspace{0.3cm}
\end{center}

Figure 17. The figure shows the graphs for $f(r)$ vs $r$ for the ultracold black hole.  Here $Q= 1.73205$ , $ M = 1.5$ and $\alpha = 0.1111$. \\

\noindent
In this case,
\begin{equation}
f(\rho)=0; \hspace{0.4 cm} f'(r)\vert_{\rho} = 0
\end{equation}
Hence the Hawking temperature, $ T_{H}(\rho)=0$.\\

To understand the geometry near the horizon, a new coordinate is defined as $ y = \eta \sqrt{ f''(\rho)/2}$. Since for the ultracold black hole $f''(\rho)=0$, one can substitute the new coordinate and take the limit $f''(\rho) \rightarrow 0$ which leads to,
\begin{equation}
ds^2 =  - \eta^2 d \psi^2 +  d \eta^2 + \rho^2 d \Omega^2
\end{equation}
The above geometry has the topology, $ R^2 \times S^2$. This is similar to the topology of the  ultracold Reissner-Nordstrom-de Sitter black hole near the degenerate horizon.

\section{Lukewarm black holes with the quintessence}

Lukewarm black hole is described as the charged-quintessence black hole  solution describing a black hole with the same temperature at an outer horizon (or event horizon) $a$ and the cosmological horizon $b$. Hence,
\begin{equation}
f(a) = f(b)=0; \hspace{0.4 cm} f'(a) = \pm f'(b)
\end{equation}
The minus sign is chosen due to the nature of the graph at $a$ and $b$. The equation(59) can be solved to obtain,
\begin{equation}
M = \frac{ a b ( a + b)}{ a^2 + 3 a b + b^2}; \hspace{0.4 cm} Q^2 = 
\frac{ a^2 b^2}{ a^2 + 3 a b + b^2}
\end{equation}
The function $f(r)$ can be re-written in terms of $a$ and $b$ as,
\begin{equation}
f(r) = \frac{ \left( 1 - \frac{a}{r} \right) \left( 1 - \frac{ b}{r} \right)
\left( a b - ( a + b) r \right)}{ (a^2 + 3 a b + b^2)}
\end{equation}
The third root corresponding to the inner horizon of the black hole is at,
\begin{equation}
r_+ = \frac {  a b}{ ( a + b) }
\end{equation}
Now, the temperature at $ a $ and $b$ is,
\begin{equation}
T(a) = T(b) = \frac{ 1 }{ 4 \pi} |f'(a)| = \frac{ 1 }{ 4 \pi} 
\frac{ | b - a|}{ (a ^2 + 3 a b + b^2)}
\end{equation}
\begin{center}
\scalebox{.9}{\includegraphics{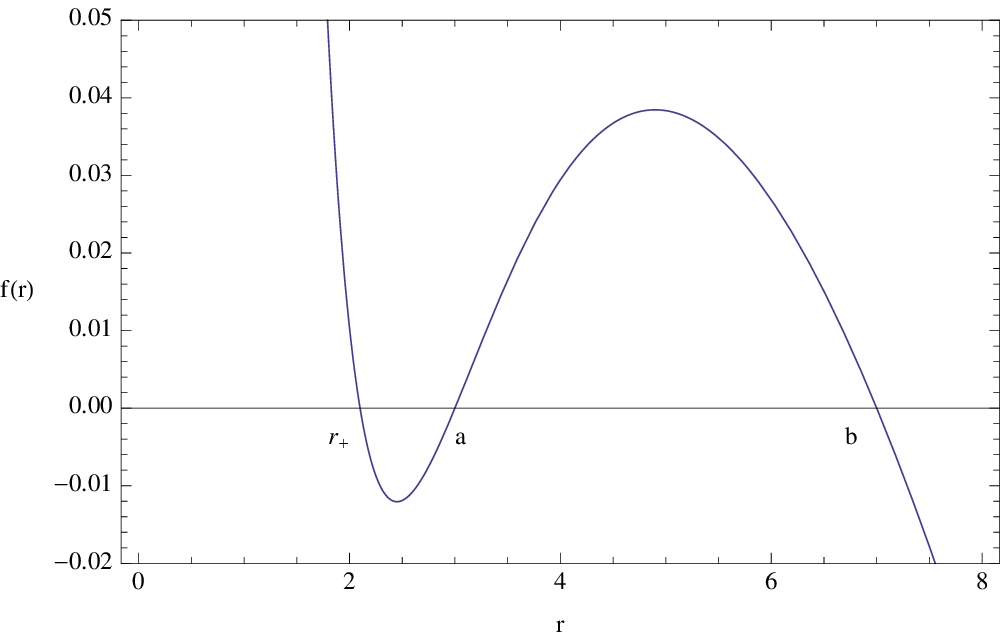}}

\vspace{0.3cm}
\end{center}

Figure 18. The figure shows the graphs $f(r)$ vs $r$ for the lukewarm Reissner-Nordstrom black hole surrounded by the quintessence. Here $a = 3$ and $ b = 7$.\\

\noindent
Lukewarm black holes for the Reissner-Nordstrom black holes are discussed in \cite{romans}. Lukewarm black holes for the quadratic gravity is presented in \cite{jerzy}.

\section{Conclusions}

We have investigated solution space for the charged black hole surrounded by the quintessence. The mass $M$ and the charge $Q$ are varied to obatin various configurations for the solutions for the horizon radii. Depending on the values of $M$, $Q$ and $\alpha$, it is possible to have three or two horizons or a single horizon.

When the horizons coincide, interesting class of black hole space-times emerges. When the Cauchy and the event horizons coincide, a cold black hole with a zero temperature emerges. The topology of this space time near the horizon is$AdS_2 \times S^2$. When the cosmological and the event horizon coincide, charged Nariai type black hole emerges. The topology at the near horizon is $ dS_2 \times S^2$. An ultra-cold black hole is a result of all three horizons converging to one. Near the horizon, this black hole has  the topology $R^2\times S^2$.  Another black hole named the lukewarm black hole emerge when the temperature of the event horizon and the cosmological horizon are the same. All these configurations are similar to what is studied in the Reissner-Nordstrom-de Sitter black hole. 

One can also study the topology of the Nariai, cold and ultracold black holes with the quintessence with general $w_q$ values (with $ -1 < w_q < -1/3$). If degenerate horizons exists, they are at $ \tilde{\rho} = \frac{ Q^2 ( w_q -1)}{ 2 M w_q}$. The topology of the Nariai, cold and ultracold black holes will be the same as for the one with $w_q  = -2/3$ but with curvature of $dS_2$ and $AdS_2$ given by $ \pm |f''(\tilde{\rho})|/2$.

\end{document}